\documentclass[twocolumn,prl,showpacs,amsmath,amssymb]{revtex4}

\usepackage{graphicx}
\usepackage{dcolumn}
\usepackage{bm,color}

\begin{document}

\preprint{APS/}

\title{Enhanced Tensor-Force Contribution in Collision Dynamics}

\author{Yoritaka Iwata$^{1}$}
 \author{Joachim A. Maruhn$^{2}$}
 \affiliation{$^{1}$GSI Helmholtzzentrum f\"ur Schwerionenforschung, D-64291 Darmstadt, Germany}
 \affiliation{$^{2}$Institut f\"ur Theoretische Physik, Universit\"at Frankfurt, D-60325 Frankfurt, Germany}

\date{\today}

\begin{abstract} 
  The tensor and spin-orbit forces contribute essentially to the
  formation of the spin mean field, and give rise to the same
  dynamical effect, namely spin polarization.  In
  this paper, based on time-dependent density functional calculations,
  we show that the tensor force, which usually acts like a small
  correction to the spin-orbit force, becomes more important in heavy-ion reactions and the effect increases with the mass of the system.
\end{abstract}


\maketitle
\section{Introduction}
The tensor force, which is necessary to explain the properties of
the deuteron, attracts special attention recently, because it has turned
out to play an essential role in the existence limit of exotic nuclei,
as well as the nuclear shell structure far from the $\beta$-stability
line (for example, see
\cite{Otsuka05,Otsuka06,Otsuka10,Brown,Lesinski,Colo,Bender,Nakamura}).
An important feature is that the spatial average of the
tensor force is exactly equal to zero, so that its effect 
is spatially localized.  On the other hand, the
spin-orbit force, which is necessary to explain the large spin
polarizations of scattered nucleons, plays a crucial role in the
nuclear shell structure. The origin of the tensor force
can be found in the one-pion exchange potential, and that of the
spin-orbit force in the relativistic aspects of quantum dynamics.

Thus the tensor and spin-orbit forces are quite different in their
origins, while resulting in the same dynamical effect, namely, spin
polarization. Spin polarization, which arises mostly from the
spin-orbit force, spontaneously takes place in the early stage of
heavy-ion reactions, and affects the equilibration process to a large
extent. As long as the microscopic time-dependent
Skyrme energy density functional (Skyrme-EDF) calculations are
concerned, the appearance of spontaneous spin polarization even in
central collisions between $\beta$-stable nuclei was shown, and its
origin was clarified to be the time-odd part of the spin-orbit force
\cite{maruhn06}.  Therefore, the enhancement or reduction of spin
polarization gives an ideal framework to pin down the properties of the
tensor force in collision situations.

In this paper, the role of the tensor force in heavy-ion reactions is
investigated based on time-dependent density functional calculations
with explicitly implemented tensor force, where the time-odd part of
the spin-orbit force is also fully taken into account. Special
attention is paid to the effect of the tensor force on time evolution.
As a result, some information on the importance of the contribution
from the tensor force in heavy-ion reactions is presented.

\section{Theoretical framework}

\begin{figure*}
\begin{center}
\includegraphics[width=40pc]{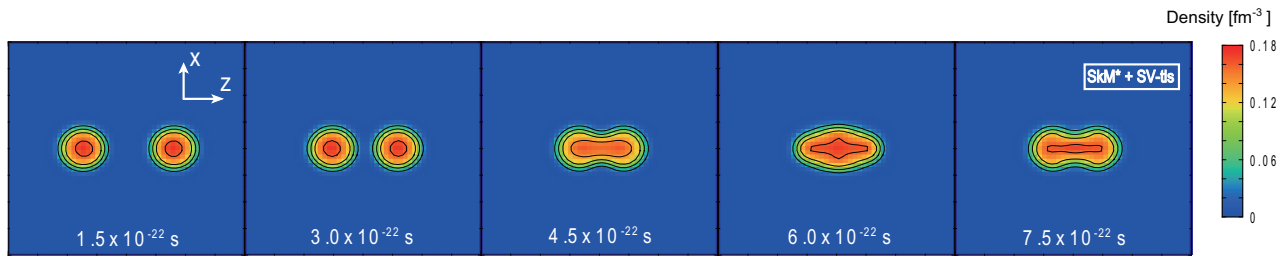}\hspace{2.5pc}%
\caption{\label{fig-1} (color online) Time evolution of $^{40}$Ca +
  $^{40}$Ca at the bombarding energy 130 MeV (c.m.).  Snapshots of the
  density are shown in a fixed square (40 $\times$ 40 fm$^{2}$) on the
  reaction plane, where contour lines are plotted for multiples of
  0.04 fm$^{-3}$.  The force used is SkM* + SV-tls. }
\end{center}
\end{figure*}

\subsection{Mean field due to spin-orbit and tensor forces}
The contribution of the tensor force, whose role was underestimated
and mostly neglected for a long time, was substantially studied in the
context of Skyrme-EDF only recently \cite{Brown,Lesinski,Colo,Bender}.
Here we begin with the functional form of the tensor and spin-orbit
forces in Skyrme-EDF.  Let $\rho$, {\boldmath $\sigma$} and ${\bf J}$
represent the number density, spin density, and spin-orbit density,
respectively.  The contribution of the tensor and spin-orbit forces to
the energy density functional has the form
\begin{equation} \label{form1} 
{\bf W}_q({\bf r}) \cdot (-i) (\nabla \times \boldsymbol{\sigma}) 
\end{equation}
where $q = n, p$ ($n$ and $p$ stand for neutron and proton,
respectively).  ${\bf W}_q({\bf r})$, which is
called the form factor of the spin mean field, is decomposed into the
contributions from spin-orbit and tensor forces.
\begin{equation}
{\bf W}_q({\bf r})  =  {\bf W}^{LS}_q({\bf r}) + {\bf W}^{T}_q({\bf r}), 
\end{equation}
where ${\bf W}^{LS}_q({\bf r})$ and ${\bf W}^{T}_q({\bf r})$ denote
the form factors of spin-orbit and tensor mean fields, respectively.
The contribution of the spin-orbit force to the functional \cite{72Vautherin} is
represented by
\[ \begin{array}{ll}
  {\bf W}^{LS}_q({\bf r}) = \frac{1}{2} W_0 (\nabla \rho({\bf r}) + 
\nabla \rho_q({\bf r})) + \frac{1}{8} (t_1 - t_2) {\bf J}_q ({\bf r}).
\end{array} \]
Note that the second term on the right-hand side,
whose contribution in collision situations was
discussed in \cite{06Umar}, has never been taken into account for
some modern Skyrme parameterizations such as SLy4d and SKM*, because it
makes fitting spin-orbit splittings more difficult.  Although there are several
versions of the tensor force, we are concerned with the natural tensor
force only.  Its contribution to the energy functional is represented by
\begin{equation} \label{wt}
  {\bf W}^{T}_q({\bf r}) = \alpha {\bf J}_q({\bf r}) 
+ \beta {\bf J}_{q'}({\bf r})
\end{equation}
with $q' = n, p$ satisfying $q \ne q'$, according to
Stancu-Brink-Flocard \cite{Stancu}, where the unique contribution of
the tensor force can be found in ${\bf J}_{q'}({\bf r})$ to ${\bf
  W}_q({\bf r})$.  The full introduction of the tensor force requires
to refit additionally the corresponding central-force parameters.
Although the full introduction brings about largely different and complicated
contributions depending on the choice of force parameter sets \cite{Stevenson},
it has been shown to mostly result in weakening the natural
contribution \cite{Reinhard}.  Several versions of the tensor force are compared in \cite{Reinhard}.
Here we restrict discussion to the tensor force as defined by Eq.~(\ref{wt}), because the
aim is not a discussion of the existence limit of exotic nuclei, but
rather the general features of the tensor contribution in reactions.
It is readily seen that the effect of the tensor force corresponds to
a quantitative modification of that due to the spin-orbit force.
Accordingly, the contribution of the tensor force should be discussed in
association with the spin-orbit force.

\subsection{Tensor-force contribution in collision situations}
A framework for measuring the effects of the tensor force is
presented with a focus on collision dynamics. Concerning
the spin polarization, it is reasonable to begin with a discussion of
spin-orbit coupling.  It is defined by the scalar triple product
\begin{equation} \label{form2} \begin{array}{ll}
{\bf L} \cdot {\bf S} 
= -i \hbar ~({\bf r} \times {\bf p}) \cdot ( \boldsymbol{\sigma} + 
\boldsymbol{\sigma'} ) \vspace{2.5mm}\\
\quad = -i \hbar ~{\bf r} \cdot ( {\bf p} \times 
\left( \boldsymbol{\sigma} + \boldsymbol{\sigma'}) \right),
\end{array} \end{equation}
where {\boldmath $\sigma$} and {\boldmath $\sigma'$} 
denote the spins of the two nucleons. In collision situations
${\bf r} \times {\bf p}$ is related to the impact parameter.
Comparing Eqs.~(\ref{form1}) and (\ref{form2}), $ {\bf W}_q({\bf r})$
in Eq.~(\ref{form1}) plays the role of the vector ${\bf r}$ in 
Eq.~(\ref{form2}), where the momentum ${\bf p}$ is 
replaced approximately by $\nabla$ in the Skyrme-EDF. 

In order to evaluate the contribution of the tensor force to
spontaneous spin polarization, we introduce a proper theoretical
setting of heavy-ion collisions.  Our starting point is that the
tensor and the spin-orbit forces are localized effects, which are not
easy to compare in collision dynamics, if there is some similarity in their localized
patterns.  Let the reaction plane be $(x, z)$ with the initial
collision direction $z$, and the direction perpendicular to the
reaction plane be $y$.  For simplicity, the spin direction of the
initial state is assumed to be parallel to the $y$-axis.  In this
setting, because only the $z$-component of ${\bf p}$ and the
$y$-component of {\boldmath $\sigma$} are non-zero, we have
\begin{equation} \begin{array}{ll} \label{form3}
 {\bf L} \cdot {\bf S} 
= -i \hbar ~ x \left( p_y \left( \sigma + \sigma' \right)_z -  p_z \left( \sigma + \sigma' \right)_y \right) \vspace{2.5mm}\\
\quad =  i \hbar ~ x   p_z \left( \sigma + \sigma' \right)_y.
\end{array} \end{equation}
We see that only the $x$-component of the vector ${\bf r}$, and thus
the $x$-component of ${\bf W}_q({\bf r})$ play a role.  In this
setting, the role of the tensor force in the spin polarization can be
evaluated by the corresponding $x$-component of ${\bf W}_q^{T}({\bf
  r})$.  Accordingly, the tensor and spin-orbit forces can be
compared, if there is a certain similarity between the $x$-components of
${\bf W}_q^{T}({\bf r})$ and ${\bf W}_q^{LS}({\bf r})$ (otherwise
attraction or repulsion happen irregularly from place to place). Note
that their similarity, which will be shown to be true, is not trivial.
In the following the $x$-components of ${\bf     W}_q^{T}({\bf r})$ 
and ${\bf W}_q^{LS}({\bf r})$ are simply represented by 
${\bf W}_q^{T}({\bf r})$ and ${\bf W}_q^{LS}({\bf r})$, 
if there is no ambiguity.

\section{Role of the tensor force}

\subsection{Spontaneous spin polarization}

A systematic three-dimensional time-dependent density functional
calculation is carried out in a spatial box $48 \times 48 \times 48$
fm$^3$ with a spatial grid spacing of 0.8 fm, in which the Skyrme-force
parameter set SV-tls \cite{Reinhard} is used for the tensor part, and
SkM* and SLy4d \cite{Chabanat-Bonche} for the remainder including the
spin-orbit force: $\alpha = 71.102$ [MeV fm$^{-5}$] and $\beta =
35.142$ [MeV fm$^{-5}$].  The parameter set SV-tls was lately introduced
in the context of the refitted tensor force; it is one of the most
reliable parameter sets in terms of reproducing the contribution of
the form factor ${\bf W}^{T}_q({\bf r})$ of the tensor mean field. 
The relative velocity in the collisions
is set to 10 \% of the speed of light, and the initial distance
of the colliding nuclei to 20.0 fm; their initial positions are
(0,0,10) and (0,0,-10).  In order to pay special attention to the
mass-dependent general features, we consider central collisions between
identical $N$=$Z$ nuclei: $^{16}$O + $^{16}$O, $^{40}$Ca + $^{40}$Ca
and $^{56}$Ni + $^{56}$Ni. The contributions from
$J_q$ and $J_{q'}$ in Eq.~(\ref{wt}) are not so different for
collisions between $N$=$Z$ nuclei, therefore the parameter
dependence mostly arises from the sum of $\alpha$ and $\beta$.
Some features of the tensor force acting on $N=Z$ bound nuclei were
studied in \cite{Bender}.

\begin{figure}
\begin{center}
\includegraphics[width=20pc]{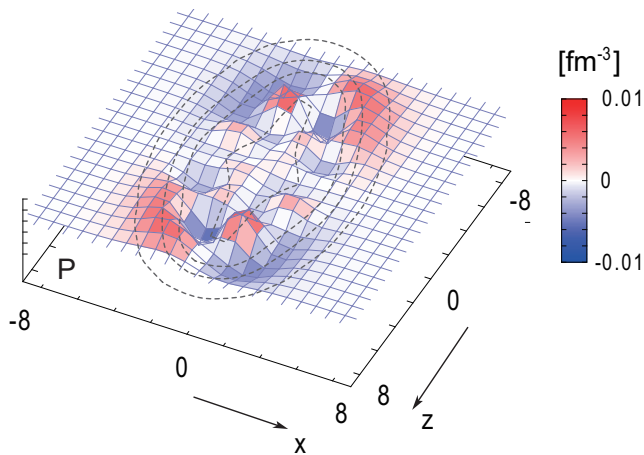}
\caption{\label{fig-2} (color online) Spin distribution (the spin is
projected onto the $y$-axis) of a compound nucleus.  A snapshot of a
composite nucleus, which corresponds to the case at time = 6.0
$\times$ 10$^{-22}$ s in Fig.~\ref{fig-1}, is shown on the reaction
plane.  For reference, contours of the density distribution are also
shown (contour = 0.01, 0.06, 0.11 and 0.16 fm$^{-3}$).}
\end{center}
\end{figure}

Figure~\ref{fig-1} shows the time evolution of $^{40}$Ca + $^{40}$Ca
resulting in fusion, where the terms associated with the tensor force
(SV-tls) are explicitly included.
The same calculation
without the spin-orbit force does not achieve fusion.  
Omitting the tensor force while including the spin-orbit force shows no notable
difference to the density evolution with all force terms
included. This suggests that large dissipation arises
from the spin-orbit force, while the tensor-force contribution is
definitely small. The composite nucleus evolves with a
continuing oscillation; the two nuclei get into contact around 
time = 4.2 $\times$ 10$^{-22}$ s, and the first full-overlap is 
achieved at 5.6 $\times$ 10$^{-22}$ s.

Let us consider the $y$-projection of spin for each single nucleon.
The spin distribution of the colliding nuclei is calculated by their superposition:
\[ P(t,{\bf r}) = \rho(t,{\bf r})_{\uparrow} - \rho(t,{\bf
  r})_{\downarrow}, \] 
where $\rho(t,{\bf r})_{\uparrow}$ and
$\rho(t,{\bf r})_{\downarrow}$ denote the densities of spin-up and
spin-down components, respectively.  In this definition, the density
plays the role of weight.  The value of $P(t,x)$ is positive if the
spin-up component is more abundant, zero for saturated spins, and
negative otherwise.  As is seen
from the presence of $\left( \sigma + \sigma' \right)_y$
in Eq.~(\ref{form3}), the problem of comparing the different role of
tensor and spin-orbit forces becomes meaningless if spontaneous spin
polarization is absent. Spin polarization appears
for all the reactions and all the force parameter sets used; e.g., in Fig.~\ref{fig-2}, the presence of spin polarization is shown for $^{40}$Ca
+ $^{40}$Ca.  As a result, the concept of examining the role of the 
tensor force in the presence of spin polarization is valid and will be
carried out in the following.

\begin{figure}
\begin{center}
\includegraphics[width=20pc]{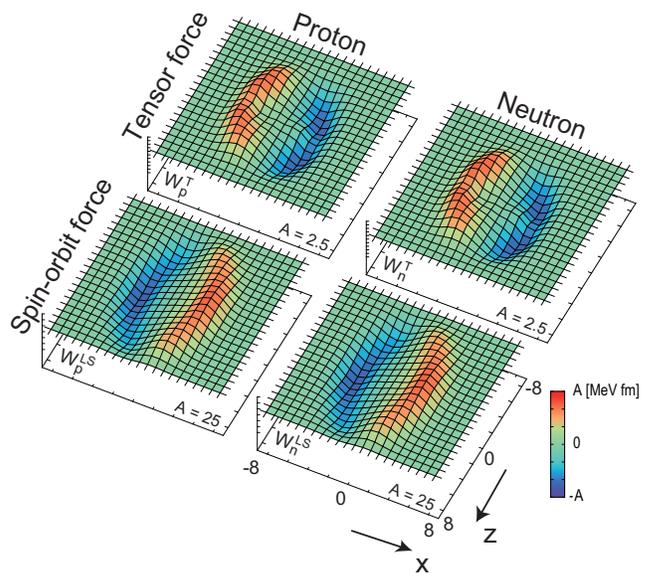}\hspace{2.5pc}%
\caption{\label{fig-3} (color online) Snapshots of the $x$-component
  of ${\bf W}_q({\bf r})$ projected on the reaction plane,
  corresponding to the case at time = 6.0 $\times$ 10$^{-22}$ s in Fig.~\ref{fig-1}.  The values are plotted separately for the tensor
  and spin-orbit forces, and for protons ($q=p$) and neutrons ($q=n$),
  respectively.}
\end{center}
\end{figure}

Figure~\ref{fig-2} shows that strong spin polarization is located on
the edge of the density distribution.  The localized pattern of the
spin structure is complicated, leading to a complicated localization
of attraction and repulsion due to the tensor force. The spin
distribution is point-symmetric with respect to the origin, which
reflects the symmetry of the central collision.  Note that the
spatial average of spin polarization for the spin-saturated system is
equal to zero.

\subsection{Comparison between tensor and spin-orbit forces}

Let us begin with the effect of the tensor force in a compound nucleus
formed briefly after the full-overlap situation (time = 6.0 $\times$
10$^{-22}$ s).  In case of $^{40}$Ca + $^{40}$Ca, Fig.~\ref{fig-3}
compares the $x$-components of ${\bf W}_q({\bf r})$ for the tensor and
spin-orbit forces.  Both distributions are antisymmetric with respect to
the $z$-axis, and have similar distributions but different signs
and amplitudes.  It is clearly seen that the tensor-force contribution
is opposite to the spin-orbit force contribution, and amounts to less than
10 percent of latter. It follows that the total
contribution from tensor and spin-orbit force is not so different from
the contribution of the spin-orbit force alone.  No significant
difference is noticed between the values for protons and neutrons.
This is expected for a collision between $N=Z$ nuclei.

\begin{figure}
\includegraphics[width=20pc]{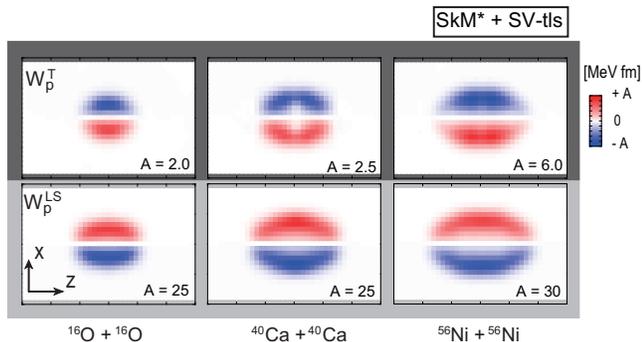}
\caption{\label{fig-4} (color online) Snapshots of the $x$-components
  of the form factors of spin mean field ${\bf W}_p^{T}({\bf r})$
  (upper ones) and ${\bf W}_p^{LS}({\bf r})$ (lower ones) at time =
  6.0 $\times$ 10$^{-22}$ s are shown in a square (30 $\times$ 20
  fm$^{2}$) on the reaction plane. The maximum amplitude $A$ of
  the function is shown in the lower right-hand side of each plot.  }
 \end{figure}

This difference in sign and the smallness of the tensor force
contribution compared to the spin-orbit force contribution is found to
hold regardless of the choice of force parameter set and the mass of
the colliding nuclei.  This difference in sign, however, did not appear
for the $y$ and $z$-components. 
On the other hand, comparing the $x$-components of ${\bf W}_q^{T}({\bf
  r})$ and ${\bf W}_q^{LS}({\bf r})$ at time = 6.0 $\times$ 10$^{-22}$
s for $^{16}$O + $^{16}$O, $^{40}$Ca + $^{40}$Ca and $^{56}$Ni +
$^{56}$Ni as shown in Fig.~\ref{fig-4}, there is a highly noticeable
increase with mass for the tensor-force contributions, while it is
only modest for those of the spin-orbit force.

\begin{figure*}
\begin{center} \qquad
\includegraphics[width=28pc]{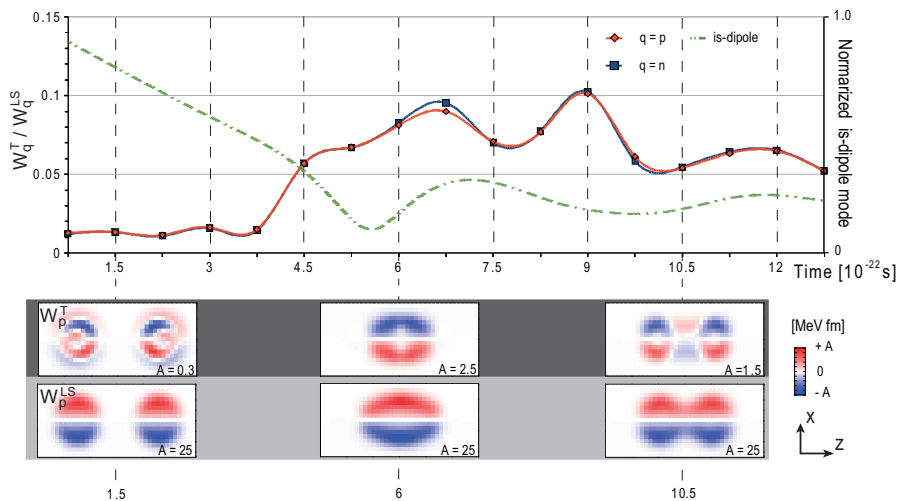}\hspace{2.5pc}%
\caption{\label{fig-5} (color online) The time evolution of the ratio
  of contributions from tensor force to those of the spin-orbit force
  is shown for protons and neutrons, respectively (upper panel), where
  the calculated points (at multiples 0.75 $\times$ 10$^{-22}$ s) are
  connected by 3rd-order spline functions.  This reaction corresponds
  to the case shown in Fig~\ref{fig-1}.  For reference, the time
  evolution of the isoscalar dipole (is-dipole) mode is shown by a
  dotted line (upper panel). The corresponding reaction-plane
  snapshots of the $x$-components of ${\bf W}_q^{T}({\bf r})$ and
  ${\bf W}_q^{LS}({\bf r})$ are shown in a square (30 $\times$ 15
  fm$^{2}$) (lower panel), where the maximum amplitude $A$ of the
  function is shown in the lower right hand side of each plot.}
\end{center}
\end{figure*}

Let us move on to the time-dependent features of the tensor force.
For the reaction shown in Fig~\ref{fig-1}, the time evolution of the ratio between tensor and spin-orbit contributions
 \begin{equation} \label{eq-ratio} {\bf W}_q^{T}/{\bf W}_q^{LS} (t) =
   \frac{ \max_{\bf r} ({\bf W}^{T}_q(t,{\bf r})) }{ \max_{\bf r} (
     {\bf W}^{LS}_q(t,{\bf r}))}
\end{equation}
is shown in Fig.~\ref{fig-5}.  In addition, the corresponding
$x$-components of ${\bf W}_p^{T}({\bf r})$ and ${\bf W}_p^{LS}({\bf
  r})$ are also shown at times 1.5 $\times$ 10$^{-22}$ s, 6 $\times$
10$^{-22}$ s and 15 $\times$ 10$^{-22}$ s.  The isoscalar dipole mode
shown in Fig.~\ref{fig-5} suggest that the full-overlap is achieved at
time = 5.5 $\times$ 10$^{-22}$ s, and the maximal elongation of
the composite nucleus at time = 7.25 $\times$ 10$^{-22}$ s.
The relaxation of the tensor contribution is
not strongly correlated  with that of the isoscalar dipole oscillation
(density oscillation towards the fused system).  The contribution
of the tensor force is quite small before the contact time (4.2
$\times$ 10$^{-22}$ s), increases after the contact time, achieves
local-maximum at times 6.75 $\times$ 10$^{-22}$ s and 9.00 $\times$
10$^{-22}$ s, and relaxes afterwards. 

Several points should be remarked here. First, the tensor-force
contribution is enhanced in collision situations, being up to 10 times
larger than before the contact time. Second, the opposite sign and the
smallness of the tensor compared to the spin-orbit contributions are
apparent during the heavy-ion collision but not before contact.  The
opposite sign means that the contribution of the tensor force
continues to weaken the spin polarization during the reaction. Third,
the similarity between protons and neutrons is confirmed throughout
the reaction.

The tensor-force contribution is compared for different force parameter sets
in Table {\ref{table2}}, where the enhancement is calculated by the
ratio
\begin{equation} \label{eq-ratio2}
 \frac {{\bf W}_q^{T}/{\bf W}_q^{LS} (t = 6.5 \times 10^{-22} s)} {{\bf W}_q^{T}/{\bf W}_q^{LS} (t = 1.5 \times 10^{-22} s)}. 
\end{equation}
where ${\bf W}_q^{T}/{\bf W}_q^{LS} (t)$ is calculated as shown in Eq.~(\ref{eq-ratio}).  This table shows that the enhancement is true
independent of the choice of force parameter sets, and no significant difference
exists between protons and neutrons.

\begin{table}  
\begin{center}
  \caption{ Enhancement of the tensor-force contribution for $^{40}$Ca
    + $^{40}$Ca.  Values of Eq.~(\ref{eq-ratio2}) are calculated for
    different force parameter sets and isospins.}
\vspace{2.5mm}
\begin{tabular}{|c|r|r|r|r|r|} \hline Parameter set & Protons ($q = p$) & Neutrons ($q = n$)  \\ 
\hline SkM* + SV-tls   & 5.94 &  6.23\\
\hline SLy4d + SV-tls   & 6.54 &  7.11\\
\hline
\end{tabular}  \label{table2}
\end{center}
\end{table}

\subsection{Mass dependence}

As already mentioned,
the opposite sign and the smallness of the tensor compared to the spin-orbit
force contribution is valid independent of the mass. Note that a
calculation using SLy4d + SV-tls showed the same features, hinting
that this is probably not strongly force-dependent.

\begin{figure}[t]
\includegraphics[width=23pc]{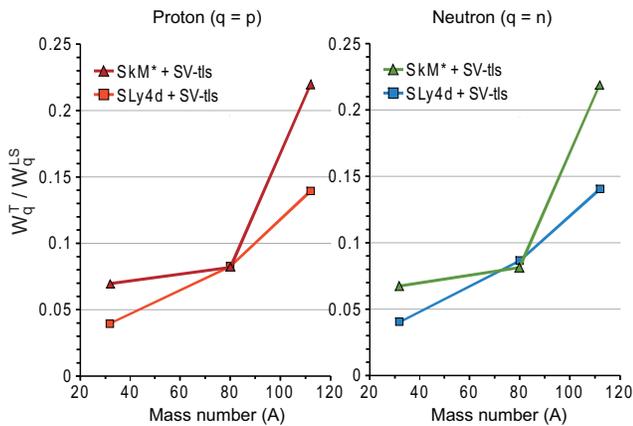}
\caption{\label{fig-6} (color online) The ratios between tensor and
  spin-orbit force contributions for protons (left panel; $q = p$) and
  neutrons (right panel; $q = n$) as functions of the mass of the
  composite nucleus.  The values at time = 6.0 $\times$ 10$^{-22}$ s
  are chosen to calculate the ratio.}
 \end{figure}

Figure~\ref{fig-6} shows the mass dependence of the ratio of tensor to
spin-orbit contributions for protons and neutron (Eq.~(\ref{eq-ratio})), where the values at time = 6.0 $\times$ 10$^{-22}$
s are chosen to calculate the ratio. In all cases, this time
corresponds to the time briefly after the first full overlap and shows
a relatively large tensor contribution close to the first maximum, so
that it is legitimate to compare the magnitude for the three cases. 
While the values are not exactly the same for the two parameter sets, 
they show the same trend; the tensor force
contribution becomes larger for reactions involving a heavier
nucleus.  For the heavier cases, 20 percent contribution from the
tensor force compared to the spin-orbit contribution is noticed (SkM*
+ SV-tls). This is
not a negligible effect considering the remarkable spin-orbit splitting
in the ground states of heavy nuclei. This
should have a certain impact on superheavy synthesis; the tensor
force is suggested to play a considerable role in whether a heavy
composite nucleus is formed successfully or not. On the other hand,
the spin polarization becomes smaller for reactions involving heavier
nuclei. The statistical ratio of spin polarization
\[ \frac { \max_{{\bf r}}({\rho_{\uparrow}(t,{\bf r}) -
    \rho_{\downarrow}(t,{\bf r})}) } { \sum_{\bf r}
  ({\rho_{\uparrow}(t,{\bf r}) + \rho_{\downarrow}(t,{\bf r})}) } \]
between time = 6.0 $\times$ 10$^{-22}$ s and 1.5 $\times$ 10$^{-22}$
s, which corresponds to the amplitude of spin polarization due to the
collision, is summarized in Table \ref{table3}. Thus the tensor-force contribution tends to survive for the heavier cases, while
the spin-orbit force contribution decreases sharply with mass.  
Note that there is no serious discrepancy between neutrons and protons
visible in Fig.~\ref{fig-6}.

\begin{table}  
\begin{center}
  \caption{Mass dependence of the growth of spin polarization (for an
    explanation see text).  For both parameter sets, values are
    normalized by the values obtained for $^{16}$O + $^{16}$O.}
\vspace{2.5mm}
\begin{tabular}{|c|r|r|r|r|r|} \hline Parameter set & $^{16}$O + $^{16}$O & $^{40}$Ca + $^{40}$Ca & $^{56}$Ni + $^{56}$Ni  \\ 
\hline SkM* + SV-tls  & 1.000 & 0.396 &  0.260\\
\hline SLy4d + SV-tls  & 1.000 & 0.571 &  0.085\\
\hline
\end{tabular}  \label{table3}
\end{center}
\end{table} 

Finally, the validity of the obtained results is also confirmed by
additionally examining  an old tensor force parameter set proposed by
Stancu-Sprung \cite{Stancu,Sprung} ($\alpha = 154.390$ [MeV fm$^{-5}$]
and $\beta = 139.910$ [MeV fm$^{-5}$]). The major difference is that
its amplitude is actually smaller than the
spin-orbit force contribution, but reaches as much as 50\% of
the spin-orbit contribution in $^{56}$Ni + $^{56}$Ni.  The difference
between the two parameters can be related to the
largeness of the $\alpha$ and $\beta$ values proposed in the Refs.
\cite{Stancu,Sprung} compared to Ref.~\cite{Reinhard}.

\section{Conclusion}
Based on time-dependent density functional calculations with
explicitly implemented tensor force, the role of the tensor force has
been studied in the context of collision dynamics.  It is remarkable
that the contribution from the tensor force is enhanced in collision
situations. Its contribution is mass-dependent and has considerable
influence on reactions involving a heavier nucleus.

As long as heavy-ion reactions between $N = Z$ identical nuclei are
concerned, the opposite sign and the smallness of the tensor force
contribution compared to that of the spin-orbit force has been
confirmed independent of mass. In particular, the
opposite sign means that the spin polarization, thus
the large dissipation due to the spin-orbit force, is reduced by
the tensor force. We conclude that the
tensor-force contribution is rather important in heavy-ion reactions
with respect to the magnitude of dissipation. The results
presented in this paper give a solid starting point
for future researches clarifying the role of the tensor force in
heavy-ion reactions involving exotic nuclei, where the
drastically different contribution from $J_q$ and $J_{q'}$ in Eq.~(\ref{wt}) might play a significant role.

This work was supported by the Helmholtz Alliance HA216/EMMI and by
the German BMBF under contract No. 06FY159D. The authors would like to
thank Prof. P. -G. Reinhard for valuable suggestions, and Prof. N.
Itagaki for fruitful discussion.

\end{document}